\documentstyle[12pt,epsf,epsfig]{article}

\newcommand{\be}{\begin{equation}}
\newcommand{\ee}{\end{equation}}
\newcommand{\bear}{\begin{eqnarray}}
\newcommand{\eear}{\end{eqnarray}}
\newcommand{\ba}{\begin{array}}
\newcommand{\ea}{\end{array}}

\newcommand{\gae}{\begin{array}{c}\,\sim\vspace{-28pt}\\>
\end{array}}

\newdimen\tdim
\tdim=\unitlength
\def\bar{\overline}

\begin{document}

\pagestyle{empty}
\begin{titlepage}
\def\thepage {}    

\title{  \bf GUT Breaking on the Lattice } 
\author{
\bf Hsin-Chia Cheng$^1$ \\[2mm]
\bf Konstantin T. Matchev$^2$ \\[2mm] 
\bf Jing Wang$^3$ \\ [2mm]
{\small {\it $^1$Enrico Fermi Institute, University of Chicago,}}\\
{\small {\it  Chicago, IL 60637, USA}}\\
{\small {\it $^2$ Theory Division, CERN}}\\ 
{\small {\it CH-1211, Geneva 23, Switzerland}}\\
{\small {\it $^3$Fermi National Accelerator Laboratory}}\\
{\small {\it P.O. Box 500, Batavia, Illinois 60510, USA}}
\thanks{E-mail: hcheng@theory.uchicago.edu, Konstantin.Matchev@cern.ch, jingw@fnal.gov.}
}


\maketitle

\vspace*{-12.0cm}
\noindent



\vbox{\baselineskip=12pt
\rightline{FERMILAB-Pub-01/223-T}
\vskip0.2truecm
\rightline{EFI-01-33}
\vskip0.2truecm
\rightline{hep-ph/0107268}
}

\vspace*{12.1cm}
\baselineskip=18pt

\begin{abstract}

  {\normalsize
We construct a supersymmetric grand unified model in the framework 
of a latticized extra dimension. 
The SU(5) symmetries on the lattice are broken by the vacuum 
expectation values of the link fields connecting adjacent SU(5) sites,
leaving just the MSSM at low energies. Below the SU(5) breaking scale, 
the theory gives rise to a similar spectrum as in orbifold breaking 
of SU(5) symmetry in 5 dimensions, and shares many features
with the latter scenario. We discuss gauge coupling unification and
proton decay emphasizing the differences with respect to the 
usual grand unified theories. Our model may be viewed as an effective 
four dimensional description of the orbifold symmetry breaking in higher 
dimensions.
} 
\end{abstract}

\vfill
\end{titlepage}

\baselineskip=18pt
\renewcommand{\arraystretch}{1.5}
\pagestyle{plain}
\setcounter{page}{1}


\section{Introduction}

Recently, a new approach to gauge theories in extra dimensions has
been introduced by considering extra dimensions on a transverse
lattice~\cite{ACG,HPW}. This provides an ``ultraviolet complete''
gauge invariant description of the higher dimensional gauge theory.
On the other hand, from a purely 4-dimensional point of view, the
extra dimensions are ``generated'' through a series of gauge groups
and link fields among them. This latticizing or desconstructing approach
to the extra dimensions provides a great tool to understand higher
dimensional gauge theories, and to obtain new models both in pure
4 dimensions and higher dimensions~\cite{Arkani-Hamed:2001nc,Cheng:2001nh,cegk,Cheng:2001an}.

In this note, we examine the orbifold breaking of the grand unified
(GUT) gauge symmetry~\cite{Kawamura:2001ev,Altarelli:2001qj,Kobakhidze:2001yk,Hall:2001pg,Hebecker:2001wq}
from the point of view of the deconstructed extra dimensions.
In the case of $SU(5)$ GUT breaking, a reflection around an orbifold
fixed point $y=0$ with the parity transformation $A_{\mu}(-y)
=P^{-1} A_{\mu}(y) P, \; P=\rm{diag}(-1,-1,-1,+1,+1)$ projects out the
zero modes of the $X,Y$ gauge bosons and breaks $SU(5)$ down
to $SU(3)\times SU(2)\times U(1)$. In field theory, the orbifold
should be viewed as a theory defined on a finite interval with suitable
boundary conditions. In this case, the boundary condition is such
that the gauge symmetry at the boundary point $y=0$ is only
$SU(3)\times SU(2)\times U(1)$, while the $SU(5)$ gauge symmetry
is preserved in the bulk. The usual gauge coupling unification can be 
preserved because the gauge couplings are dominated by the $SU(5)$
symmetric bulk contributions which are enhanced by the volume
factor relative to the contributions from the boundary.
There are several nice features of this GUT breaking mechanism.
It is easy to obtain doublet-triplet splitting in the Higgs sector,
at the same time avoiding proton decay mediated by
the colored triplet Higgs fields, which may already pose a problem with 
the current experimental bounds in the usual 4-dimensional GUT.
The gaugino mediated supersymmetry (SUSY) breaking~\cite{Kaplan:2000ac} can naturally
be incorporated in this framework to solve the SUSY flavor problem.

In the following, we consider this orbifold GUT breaking on
latticized extra dimensions. It becomes a 4-dimensional theory
with a series of gauge groups, broken down to the diagonal subgroup
by the link field vacuum expectation values (VEV's). The simplest
realization is to have only $SU(3)\times SU(2)\times U(1)$ gauge
symmetry on one lattice point at the end, and $SU(5)$'s on all other 
lattice points. However, we prefer to start with $SU(5)$'s on all lattice
points, and break them down to $SU(3)\times SU(2)\times U(1)$ with
the VEV's of the link fields. Since the link fields are identified
with the $A_5$ component of the gauge field in the continuum limit, 
this is equivalent to the ``Wilson line breaking,'' which has been
shown to be equivalent to the orbifold breaking in the continuum
theory~\cite{Hebecker:2001jb,Hosotani:1983xw,Candelas:1985en}. In this model, the 
$SU(3)$, $SU(2)$, and $U(1)$ gauge couplings are truly unified
at some high scale. We will find that the spectrum of the continuum
theory is reproduced in the limit of large number of lattice points.
We will also discuss related issues such as doublet-triplet
splitting and gauge coupling unification in this 4-dimensional picture.
While finishing this work, we learned that a similar idea is being
pursued by C.~Csaki, G.~D.~Kribs, and J.~Terning~\cite{CKT}.

\section{Formalism}

\subsection{SUSY SU(5) on the orbifold lattice}
\label{sec:susysu5}

We will begin with a supersymmetric $SU(5)$ theory on a latticized extra dimention.
We assume that we have $N+1$ $SU(5)$ gauge groups with common 
gauge coupling $g$,  with $N+1$ $N=1$ vector multiplets $V_i$ ($i=0,\cdots,N$), 
one for each $SU(5)$. 
There are also two sets of $N$ chiral multiplets $\Phi_{i}$ and $\bar{\Phi}_i$, 
$\Phi_i$ forms 
($\bar{5}_{i-1}, 5_{i}$) under the two nearest $SU(5)$'s,  while $\bar{\Phi}_i$ has the
opposite charges. The Lagrangian for the vector
multiplets and the chiral fields is the following:
\be
{\cal L} = \int d^4x \left[ \int d^4\theta \sum_{i=1}^{N}\left( \Phi_{i}^{\dagger} 
e^{(V_{i-1}-V_{i})} \Phi_{i} + \bar{\Phi}_{i}^{\dagger} e^{(-V_{i-1}+V_{i})} 
\bar{\Phi}_{i} \right) + \int d^2 \theta \sum_{i=0}^{N} W_{i}^{\alpha} W_{i,\alpha} \right]. 
\ee

As shown in \cite{cegk}, if the diagonal components of the
link fields, $\Phi_{i}$ and $\bar{\Phi}_i$, acquire
universal vacuum expectation values $v/\sqrt{2}$ which preserve
the $N=1$ supersymmetry,
\be
\langle \Phi_{i} \rangle = \langle {\bar{\Phi}_i} \rangle 
= \frac{v}{\sqrt{2}}{\rm diag}{(1,1,1,1,1)},
\ee
then $SU(5)^{N+1}$ is spontaneously
broken down to a diagonal $SU(5)$.\footnote{For simplicity and ease of comparison with
the result of orbifold breaking in 5D, we assume that the gauge couplings and the 
link VEV's are the same for all lattice points.} The vector
multiplets have the mass spectrum $M_{V,n} =
2gv\sin{\frac{n\pi}{2(N+1)}}$, $n=0 \cdots N$, while certain linear 
combinations of some components in the link fields $\Phi$ and $\bar{\Phi}$, 
which become part of the massive $N=1$ vector multiplets, receive $D$ term 
contributions and acquire the mass spectrum $M_{\Phi, \bar{\Phi},n} =
2gv\sin{\frac{n\pi}{2(N+1)}}$, $n=1 \cdots N$. The other components of $\Phi$ and 
$\bar{\Phi}$ acquire masses $\sim v$ or higher and thus decouple from the low 
energy effective theory.  Therefore, one recovers an $N=1$ $SU(5)$ theory 
at the zero mode level. 

In addition, we have four sets of chiral fields: 
$H_{5,i}=\{H_{C,i}, H_{U,i}\}$ and its conjugate $H_{5,i}^c=\{H_{C,i}^c, H_{U,i}^c\}$;
as well as $H_{\bar{5},i}=\{H_{\bar{C},i},H_{D,i}\}$ and its conjugate 
$H_{\bar{5},i}^c=\{H_{\bar{C},i}^c,H_{D,i}^c\}$, 
where the subscripts show the charges of the fields under 
each $SU(5)$. We assume that on the zeroth brane, one only 
has $H_{5,0}$ and $H_{\bar{5},0}$, but not their conjugate partners. 
The superpotentital for these fields is the following, 
\be
\label{H}
{\cal L}  \sim \int d^4x \int d^2 \theta \sum_{i=1}^{N} ( M_H H_{5,i} H_{5,i}^c 
- \lambda \Phi_{i} H_{5,i-1} H_{5,i}^c + M_H H_{\bar{5},i} H_{\bar{5},i}^c 
- \lambda \bar{\Phi}_{i} H_{\bar{5},i-1} H_{\bar{5},i}^c) ...
\ee
When $\Phi_i$ and $\bar{\Phi}_i$ accquire VEV's, 
and assuming $M_H = \lambda v/\sqrt{2}$, the mass spectra 
for the $H$ fields arising from the superpotential are such that $H_{5}$ 
and $H_{\bar{5}}$ have massless zero modes which  preserve $N=1$ SUSY, 
while all conjugate fields become massive. The massive $H$ and $H^c$ 
fields have the spectra $M_{H, H^c,n} = 2 M_H \sin{\frac{n\pi}{2(N+1)}}$, 
$n=1, \cdots, N$, which is the same as the massive vector multiplets and 
the massive chiral link fields, given the choice $ M_H = g v \; (\lambda=\sqrt{2} g)$. 

The results map onto a continuum five-dimensional theory with $N=1$ supersymmetry 
compactified on a $Z_2$ orbifold of size $L = (N+1)/gv$. 
Orbifolding breaks the $N=1$ SUSY in five dimensions (which is equivalent 
to $N=2$ SUSY in four dimensions) down to $N=1$ SUSY in four dimensions. 
The Higgs fields $H_5$ and $H_{\bar{5}}$ are complete hypermultiplets
in the 5D theory, while in 4D $N=1$ language each of them includes two chiral 
multiplets that are conjugate of each other.   

\subsection{SU(5) breaking}

To generate $SU(5)$ breaking, we assume that the first set of 
link fields takes on a different form. We assume that there are four 
link fields, $\Phi_{1}$, $\bar{\Phi}_1$, $\Phi_{1}^{'}$, $\bar{\Phi}_1^{'}$ 
that are charged under $SU(5)_0$ and $SU(5)_1$. $\Phi_{1}$ and 
$\Phi_{1}^{'}$ form ($\bar{5}_0$, $5_1$) representation, 
and $\bar{\Phi}_{1}$ and $\bar{\Phi}_{1}^{'}$ form (${5}_0$, $\bar{5}_1$).  
Their VEV's have the following structure, 
\be
\begin{array}{c}
\langle  \Phi_{1}  \rangle = \langle \bar{\Phi}_1 \rangle 
= \frac{v}{\sqrt{2}}{\rm diag}{(1,1,1,0,0)}; \\
\langle  \Phi_{1}^{'}  \rangle = \langle \bar{\Phi}_1^{'} \rangle 
= \frac{v}{\sqrt{2}}{\rm diag}{(0,0,0,1,1)}; 
\end{array}
\ee
These VEV's can be obtained with suitable superpotential interaction~\cite{BDS,S}.
All other link fields have the same structure and VEV's as previously discussed. 
The unbroken gauge group is then $SU(3) \times SU(2) \times U(1)$, 
which is easily seen from the mass spectrum of the gauge bosons. 
For the $SU(3)\times SU(2) \times U(1)$ gauge bosons, the mass matrix 
remains the same as in the case considered previously in Sec.~\ref{sec:susysu5},
\be 
M^{2}_{3-2-1}= \frac{1}{2} g^2 v^2  \left(
\begin{array}{ccccc}
1&-1&0&\cdots&0 \\
-1&2&-1& \cdots&0 \\
0&-1 &2 &\cdots&0 \\
& & & \cdots & \\
0&0&\cdots&-1&1
\end{array} \right).
\ee
Hence, there is a zero mode for each of the gauge groups. 
The $X$, $Y$ gauge bosons, however, accquire a different mass spectrum, 
due to the fact that the VEV's of $\Phi_1$ and $\Phi_1^{'}$ do not generate 
off-diagonal mass terms between the gauge bosons of $SU(5)_0$ and $SU(5)_1$,  
\be 
M^{2}_{X,Y}= \frac{1}{2} g^2 v^2  \left(
\begin{array}{ccccc}
1&0&0&\cdots&0 \\
0&2&-1& \cdots&0 \\
0&-1 &2 &\cdots&0 \\
& & & \cdots & \\
0&0&\cdots&-1&1
\end{array} \right).
\ee
As a result, the $X$, $Y$ gauge bosons on the $0$th brane are decoupled 
from the rest of the lattice, and have masses $\bar{M}_0=gv$. 
The other $X$, $Y$ bosons on branes $1..N$ accquire the mass 
spectrum $M_{X,Y,n} = 2 gv \sin{\frac{(n-1/2)\pi}{(2N+1)}}$, $n=1, \cdots N$. 

Since the model preserves $N=1$ SUSY, we expect it to contain the 
full vector multiplets of $SU(3)\times SU(2)\times U(1)$, and 
the $X$, $Y$ vector multiplets to exhibit the same mass spectrum 
as their scalar components. The corresponding components in the 
$\Phi$ and $\bar{\Phi}$ fields also split in a similar fashion.  

In the Higgs sector, we modify the couplings between the Higgs fields on 
the $0$th and $1$st brane and the corresponding link fields, 
while keeping the couplings on all other branes the same. 
The superpotential takes the following form, 
\be
W \sim \lambda^{'}\frac{H_{5,0} \Phi_{1} \bar{\Phi}_{1} H_{\bar{5},0}}{M} 
- \lambda H_{5,0} \Phi_1^{'} H_{5,1}^c + M_H H_{5,1} H_{5,1}^c  
- \lambda H_{\bar{5},0} \Phi_1^{'} H_{\bar{5},1}^c 
+ M_H H_{\bar{5},1} H_{\bar{5},1}^c + ...
\label{WH}
\ee
where the $...$ include the couplings of the $H,~ H^c$ fields 
present in Eq.~(\ref{H}). 

Since $\Phi_{1}^{'}$ and $\bar{\Phi}_{1}^{'}$ have non-zero VEV's only 
in their last two diagonal components, the Higgs doublets $H_{U,i}$ and 
$H_{D,i}$ from $H_{5,i}$ and $H_{\bar{5},i}$ accquire the same mass 
spectrum as the $SU(3) \times SU(2) \times U(1)$ vector multiplets,
as we previously discussed. 
Namely, $M_{H_U, H_D, n} = 2gv\sin{\frac{n\pi}{2(N+1)}}$, $n=0 \cdots N$.  
However,  the structure for the colored Higgs components is changed. 
$\Phi_{1}^{'}$ and $\bar{\Phi}_{1}^{'}$  do not generate off-diagonal
mass terms between the $0$th and $1$st colored Higgs field, hence, the
colored triplets on the $0$th brane $H_{C,0}$ and $H_{\bar{C},0}$ 
are decoupled from the rest of the lattice.  
The $N \times N$ mass matrix for the colored Higgses on the 
$n=1, \cdots N$ branes takes the following form, 
\be
M^2_{H_C, H_{\bar{C}}} = M_H^2 \left(
\begin{array}{ccccc}
2&-1&0&\cdots&0 \\
-1&2&-1& \cdots&0 \\
0&-1 &2 &\cdots&0 \\
& & & \cdots & \\
0&0&\cdots&-1&1
\end{array} \right).
\ee    
Therefore, there are $N$ massive modes, with the spectrum 
$M_{H_C, H_{\bar{C}},n} = 2 gv \sin{\frac{(n-1/2)\pi}{(2N+1)}}$, $n=1, \cdots N$. 
Finally, the colored components in $H_{5,0}$ and $H_{\bar{5},0}$ accquire 
masses from the higher dimensional coupling that is localized on the 
first brane, as shown in eqn.(\ref{WH}). Their masses are $\lambda^{'} v^2/2M$. 
One can tune the parameter $\lambda^{'}$, assuming that $v$ is comparable 
to $M$, such that $ \lambda^{'} v^2/2M = gv $. Hence, the complete colored
Higgs spectrum matches onto that of the $X$, $Y$ vector multiplets. 

It is easy to verify that the $H^c$ fields also exhibit the same 
splitting between their colored components and their doublet components, 
due to the vacuum structure of the first set of link fields. 
The doublet components of the $H^c$ fields have the spectrum 
$M_{H^c_{U,D}}= 2gv\sin{\frac{n\pi}{2(N+1)}}$, $n=1 \cdots N$, 
while the triplets have the spectrum 
$M_{H^c_{C}} = 2 gv \sin{\frac{(n-1/2)\pi}{(2N+1)}}$.   

In summary, the massless modes in our model include
$N=1$ $SU(3)\times SU(2) \times U(1)$ vector multiplets and
two Higgs chiral multiplets $H_U$ and $H_D$. 
The massive modes fall into two types according to their spectrum. 

\begin{itemize}

\item $M_{1n} = 2 gv \sin( \frac{n\pi}{2(N+1)} )$, $n=1, \cdots, N$. 
The fields that have this type of mass spectrum are the KK modes 
of the $SU(3) \times SU(2) \times U(1)$ gauge supermultiplets, 
which include components coming from the link fields $\Phi$ and
$\bar{\Phi}$, and the KK towers of Higgs doublets including $H_{U,D}$ 
and $H_{U,D}^c$.

\item $\bar{M}_0= gv$ and $M_{2n} = 2 gv \sin{\frac{(n-1/2)\pi}{(2N+1)}}$, 
$n=1, \cdots N$. This category includes the massive $X$, $Y$ vector 
multiplets, which contain components from the link fields, and
the KK towers of Higgs triplets which include $H_{C, \bar{C}}$. 
At the same time, the massive colored Higgs modes belonging to
the $H_{5, \bar{5}}^c$ (there is a total of $N$ of those)
do not include $\bar{M}_0$. 

\end{itemize}   

There are other components of the link fields (and other possible fields
required to generate the link VEV's) acquiring masses of order $v$ or
higher from minimizing the potential.

At $n\ll N$, $M_{1n} \approx  gv \frac{n\pi}{N}$, 
while $M_{2n} \approx gv \frac{(n - 1/2)\pi}{N}$. 
Hence, the masses of the two sets of KK modes have 
a relative shift of $\frac{gv \pi}{2N}$. The low energy 
spectrum is the same as that of the KK modes in \cite{Hall:2001pg}, 
in which a SUSY $SU(5)$ model in five dimensions is compactified 
on a $Z_2\times Z_2 $ orbifold.

One complete family of quarks and leptons comes from a $\bar{\bf 5}$ 
and a ${\bf 10}$ of the $SU(5)$. We can assume that these matter 
fields are localized on a single lattice point (i.e., transforming  
under a single gauge group). Having matter fields localized 
on the boundary which preserves (breaks) the $SU(5)$ gauge 
symmetry in the continuum theory corresponds to having them 
transforming under the $N$th ($0$th) gauge group. Alternatively,
they can have wavefunctions distributed in the latticized
bulk if one adds ${\bf 10}$, $\bar{\bf 10}$ and $\bar{\bf 5}$, 
${\bf 5}$ on several lattice points, linked by the $\Phi$, 
$\bar{\Phi}$ fields as in the Higgs sector. Because the zero
modes of the Higgs doublets are equal linear combinations of 
$H_{U,i}$ and $H_{D,i}$ on all lattice points, they couple to 
fermions localized on different branes through Yukawa 
couplings and generate masses and mixings for 
the standard model fermions after the electroweak symmetry is broken.

\section{Discussion}


Given the spectrum presented in the previous section, the running 
of the gauge couplings at the 1-loop level including the threshold 
corrections from all massive modes can be easily calculated as follows, 
\be
\alpha^{-1}_a (M_Z) = \alpha^{-1}_{G} (M_*) 
+ \frac{1}{2\pi} \left[ \beta_a \ln(\frac{M_*}{M_Z}) 
+ \gamma_a \sum_{n=1}^{N} \ln(\frac{M_*}{M_{1n}}) 
+ \delta_a \sum_{n=1}^{N} \ln(\frac{M_*}{M_{2n}}) 
+ \delta^{'}_a \ln(\frac{M_*}{\bar{M}_0}) + \Delta_a \right] \ .
\ee
Here $\alpha_G=g^2/(4\pi(N+1))$, and
the numerical coefficients are determined only by the group 
structure of the fields. $\beta_a$ ($a=1,2,3$ refers 
to $U(1)$, $SU(2)$ and $SU(3)$) includes the contribution 
from the zero modes, $\gamma_a$ includes the contribution
from the modes which have a Type I mass spectrum,  
$\delta_a$ accounts for the the modes with a Type II mass spectrum
for $n=1, \cdots N$. These coefficients have been calculated 
in \cite{Hall:2001pg}, where a model with a similar spectrum 
has been constructed from a $Z_2 \times Z_2$ compactification 
of a supersymmetric 5D theory: 
$\beta_a= (\frac{33}{5}, 1, -3)$, 
$\gamma_a= (\frac{6}{5}, -2, -6)$, 
$\delta_a= (-\frac{46}{5}, -6, -2)$.  
$\delta^{'}_{a}$ counts the contributions from the $X$ and $Y$ 
gauge bosons and $H_{C, \bar{C}}$, both with mass $\bar{M}_0$. 
It is easy to show that $\delta^{'}_a = (-\frac{48}{5}, -6, -3)$. 
$\Delta_{a}$ includes the threshold corrections from heavy 
link field components, which are near or above 
the $SU(5)$ breaking scale. 

As discussed in Refs.~\cite{Hall:2001pg,Nomura:2001mf}, gauge coupling
unification is not ruined by the presence of the Kaluza-Klein spectrum.
We now examine this in our model in more detail. 

Let us define $M_G= 2\times 10^{16}$ GeV as the scale where $\alpha_1$ and $\alpha_2$
meet in the MSSM. Previous studies \cite{Feng:2001bp,Bagger:1997ei} have shown that with the
central values for the gauge couplings at the weak scale, and a
SUSY spectrum which is not unnaturally heavy, the gauge couplings 
miss each other at the scale $M_G$ by
\begin{equation}
\varepsilon_3 \equiv {g_3-g_1\over g_1} \sim -(1-2) \% \ .
\label{eps3}
\end{equation}
This mismatch should be accounted for by the GUT-scale threshold corrections within
any specific grand unified model. We now proceed to calculate the prediction for
$\varepsilon_3$ in our model.

We choose to match the MSSM onto the full GUT theory at the scale $M_\ast = M_G$.
The condition $\alpha_1(M_G)=\alpha_2(M_G)$ implies that the threshold corrections to 
$\alpha_1$ and $\alpha_2$ at the scale $M_G$ should be equal.
This allows us to compute the value of $\bar{M}_{0} = gv$ for any given fixed $N$:
\be
\ln \frac{M_G}{\bar{M}_0} = -\frac{8}{9} (G_N-D_N),
\ee
where the numerical factors $G_N$ and $D_N$ are defined as follows, 
\bear
G_N &\equiv & \sum_{n=1}^{N} \ln \left[ 2\sin \frac{n\pi}{2(N+1)}\right]
\ = \ {1\over2}\ \ln (N+1),
\nonumber \\
D_N &\equiv & \sum_{n=1}^{N} \ln \left[ 2\sin 
\frac{(n-1/2)\pi}{2N+1}\right] \ =\ 0 \ .
\eear
(In what follows, we ignore the model-dependent effects from $\Delta_a$.)

\begin{figure}[tbp]
\epsfig{file=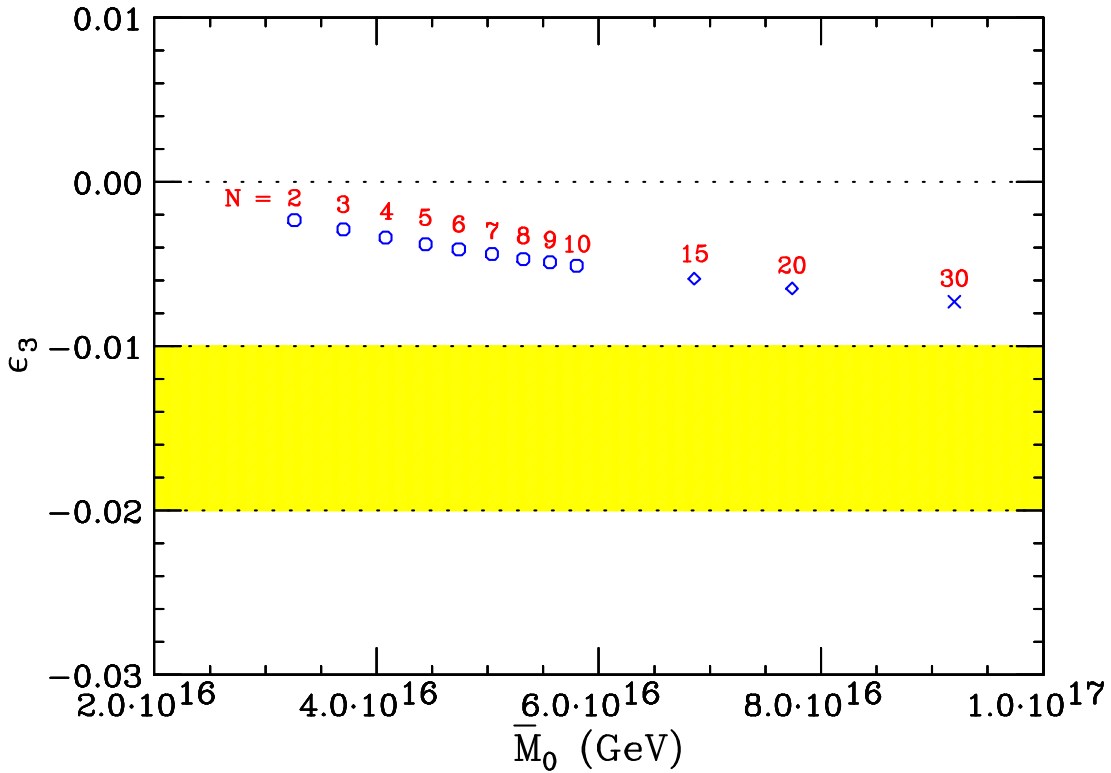,height=4in}
\caption{ Predictions for $\varepsilon_3$ 
and $\bar{M}_{0}$ for several different values of $N$.
The circles (diamonds, crosses) are consistent (marginally consistent, excluded)
with the proton decay limits from dimension 6 operators. The shaded region is the 
range of $\varepsilon_3$ preferred by low energy data (see Eq.~(\ref{eps3})). }
\label{fig:gu}
\end{figure}

Having determined $\bar{M}_{0}=gv$, there are no free parameters left, 
and for any given $N$ we get a 
prediction for $\varepsilon_3$ at the unification scale $M_G$:
\be
\varepsilon_3 = -\frac{\alpha_G}{3\pi} (G_N-D_N).
\ee
In Fig.~\ref{fig:gu} we show the prediction for $\varepsilon_3$ 
and $\bar{M}_{0}$ for several different values of $N$. 
For $N\gae 20$ the proton decay rate from the dimension 6 operator 
exceeds the experimental bound, as discussed below. The points
which are consistent with (marginally consistent with, excluded by)
proton decay, are denoted by circles (diamonds, crosses). 
We see that the predicted threshold correction $\varepsilon_3$
is {\em negative}, i.e. goes in the right direction.
However, its magnitude is not large enough to completely fix
gauge coupling unification. One might hope that 
the additional threshold effects $\Delta_a$ due to the
heavy components of the link fields will ameliorate the situation.
Alternatively, gauge coupling unification can be further 
improved by reducing $\lambda'$, hence
lowering the mass of the colored triplet Higgs 
on lattice point 0, which results in an addtitional 
negative contribution to $\varepsilon_3$.

From Fig.~\ref{fig:gu} we also see that the $SU(5)$ breaking scale, 
defined as $2gv=2\bar{M}_{0}$, 
is a few times higher than the usual $M_G$, and it grows for larger $N$. 
The mass of the lowest KK mode, namely, the effective compactification
scale, is between $0.4 \sim 0.8 \times M_G$.         
     

The colored triplet Higgs mediated proton decay is absent if the matter fields
are localized away from the zeroth lattice point (i.e., do not transform under $SU(5)_0$),
because the two sets of Higgs fields containing $H_U$ and $H_D$ do not couple
to each other away from lattice point 0. Although the triplets on 
lattice point 0, $H_{C,0}$ and $H_{\bar{C},0}$, couple
through the non-renormalizable interaction $\lambda'$, they decouple from the
triplets on the other lattice points. As a result, the proton decay 
process mediated by $H_{C,0}$ and $H_{\bar{C},0}$ can only take place
if the quarks and the leptons are on the 0th brane.\footnote{In Ref.~\cite{Hall:2001pg},
a $U(1)_R$ symmetry is imposed to completely forbid the dimension 5 proton
decay operators. This $U(1)_R$ symmetry is not respected by the non-renormalizable
interaction $\lambda'$ in our model. However, the size of the dimension 5 proton 
decay operators depends on the flavor structure \cite{Murayama:1994tc} and 
hence is difficult to estimate without a flavor theory.}

If the matter fields are localized
on branes away from the $0$th brane,  the dimension 6 proton decay 
operators from the $X,Y$ gauge boson exchange will be enhanced compared 
to the usual SUSY GUT, because there are many $X,Y$ gauge bosons 
contributing to the process and the lightest ones are lighter than 
those in the traditional 4D SUSY GUT. 
The experimental value of the proton lifetime thus imposes constraints 
on the scales in our construction. The decay mode 
$p \rightarrow e^{+} \pi^{0}$ through exchanging of $X$ and $Y$ gauge 
bosons requires that the lightest $X$ and $Y$ gauge bosons both
should have mass $gv \pi /(2N+1) \ge 5 \times 10^{15}$ GeV.
On the other hand, as we discussed ealier, 
the $X,Y$ gauge bosons on lattice point 0 are decoupled from the other 
$X,Y$ gauge bosons, and have mass $gv$, which is not supressed by the 
volume factor $N$ and  somewaht larger than the
usual SUSY GUT scale. Therefore, 
if the matter fields are localized on the lattice 0, the dimension 6 proton
decay operators will be suppressed compared to the case when 
matter is localized away from the 0th brane.


As mentioned in the Introduction, gaugino mediated SUSY breaking can be easily
incoporated in the orbifold GUT breaking scenario. In our case, similar 
superpartner spectrum can be obtained if SUSY breaking only couples to the
gauge group on the lattice point away from where matter fields are 
localized~\cite{cegk,Cheng:2001an}.

In summary, we have constructed a 4D SUSY GUT theory with many $SU(5)$ gauge
groups. The gauge symmetry breaking scale is somewhat higher than the 
GUT scale 
in the usual 4D theory. However, gauge coupling unification is achieved
due to the threshold corrections from the ``Kaluza-Klein'' modes
lighter than the symmetry breaking scale. It shares many features with
the 5D orbifold GUT breaking models, and may be viewed as an effective
4D description of these higher dimensional mechanisms.

\newpage
\noindent
{\bf \Large \bf Acknowledgements}

We wish to thank Y. Nomura
for useful discussions.
We also wish  
to acknowledge the kind hospitality of the
Aspen Center for Physics where this work was finished. 
H.-C. Cheng is supported by the Robert R. McCormick Fellowship and by   
DOE grant DE-FG02-90ER-40560.    
Research by JW was supported by the U.S.~Department of Energy   
Grant DE-AC02-76CHO3000.   

\frenchspacing
\vspace*{1cm}


\end{document}